\documentclass[sigconf,10pt]{acmart}
\usepackage{tabularx,booktabs}
\newcolumntype{Y}{>{\centering\arraybackslash}X}
\def\BibTeX{{\rm B\kern-.05em{\sc i\kern-.025em b}\kern-.08emT\kern-.1667em\lower.7ex\hbox{E}\kern-.125emX}}
    
\copyrightyear{2019}
\acmYear{2019}
\setcopyright{acmlicensed}
\acmConference[DEEM '19]{DEEM '19: Workshop on Data Management for End-to-End Machine Learning}{June 30, 2019}{Amsterdam, NL}
\acmBooktitle{DEEM workshop}
\acmPrice{}
\acmDOI{}
\acmISBN{}

\newcommand{\mc}{{\sf MLClean}}

\begin{document}

%
\title[Data Cleaning for Accurate, Fair, and Robust Models]{Data Cleaning for Accurate, Fair, and Robust Models: \\A Big Data - AI Integration Approach}


%

\author{Ki Hyun Tae, Yuji Roh, Young Hun Oh, Hyunsu Kim, Steven Euijong Whang}
\authornote{Corresponding author}
\email{{kihyun.tae,yuji.roh,oh.henry,hyunsu00,swhang}@kaist.ac.kr}
\affiliation{%
  \institution{KAIST}
}

%
\renewcommand{\shortauthors}{Tae et al.}

%
\begin{abstract}
The wide use of machine learning is fundamentally changing the software development paradigm (a.k.a. Software 2.0) where data becomes a first-class citizen, on par with code. As machine learning is used in sensitive applications, it becomes imperative that the trained model is accurate, fair, and robust to attacks. While many techniques have been proposed to improve the model training process (in-processing approach) or the trained model itself (post-processing), we argue that the most effective method is to clean the root cause of error: the data the model is trained on (pre-processing). Historically, there are at least three research communities that have been separately studying this problem: data management, machine learning (model fairness), and security. Although a significant amount of research has been done by each community, ultimately the same datasets must be preprocessed, and there is little understanding how the techniques relate to each other and can possibly be integrated. We contend that it is time to extend the notion of data cleaning for modern machine learning needs. We identify dependencies among the data preprocessing techniques and propose \mc{}, a unified data cleaning framework that integrates the techniques and helps train accurate and fair models. This work is part of a broader trend of Big data -- Artificial Intelligence (AI) integration.
\end{abstract}

%
%


%

\maketitle

\section{Introduction}
\label{sec:introduction}

Machine learning is becoming a new paradigm of programming (called Software 2.0~\cite{karpathi}) where data becomes a first-class citizen, on par with code. While existing development involved implementing certain functionalities and testing code, the development process of machine learning applications involves data collection, data analysis and validation, model training, and model validation where the entire process can repeat~\cite{Polyzotis:2017:DMC:3035918.3054782}. As an example, Google has developed and is opensourcing its end-to-end machine learning platform TensorFlow Extended (TFX)~\cite{DBLP:conf/kdd/BaylorBCFFHHIJK17}, in order to democratize AI and make machine learning development easy for anyone.

As machine learning becomes widely used even in sensitive applications like self-driving cars and medical applications, training a model needs to be more ``bullet proof'' as well. In addition to being accurate, machine learning models should not discriminate certain demographics of people or be sensitive to adversarial data. However, machine learning is only as good as its data, and no matter how good the training algorithm, the ultimate problem may lie in the data itself. 


Interestingly, multiple communities have been investigating how to prepare data for machine learning (see Section~\ref{sec:techniques} for details). Unfortunately, little is known how the different data preprocessing approaches depend on each other and can be used together when a dataset is dirty, biased, and adversarial. In addition, we cannot assume that solving one problem will automatically solve the others. From a data management standpoint, we contend that it is a good time to {\em extend the data cleaning problem} for the pressing needs of modern machine learning for accurate, fair, and robust model training.

In this paper, our novel contributions are making a comparison of three data preprocessing techniques and proposing \mc{}, a data cleaning framework that performs traditional data cleaning, unfairness mitigation, and data sanitization together. \mc{} views machine learning models as black boxes, which means it can support any model, but cannot exploit the internals of it. \mc{} is an example of the larger trend of Big data and AI integration where data management and machine learning techniques are converging and opens up many new research opportunities.

\section{Extending Data Cleaning}
\label{sec:techniques}

We compare and identify dependencies among the three data preprocessing techniques and discuss how data cleaning can possibly be extended to the other preprocessing techniques.


\subsection{Traditional Data Cleaning} 
Data cleaning~\cite{DBLP:conf/sigmod/ChuIKW16} originates from the data management community and has been studied for decades. Traditionally, there is a focus on cleaning structured data with schema at scale where integrity constraints, denial constraints, and functional dependencies need to be satisfied. In addition, duplicates must be removed, and values need to be corrected to be within certain ranges or to exist in external data sources. More recently, there are efforts to improve machine learning accuracy~\cite{DBLP:conf/sigmod/DongR18} and data validation techniques for machine learning pipelines~\cite{Polyzotis:2017:DMC:3035918.3054782}. However, these techniques do not resolve the pressing issues of model fairness or model robustness against adversarial data. 

As a running example, suppose we are cleaning a set of training examples in Table~\ref{tbl:examples} (small for illustration purposes). This data is not clean in the sense that $e_2$ and $e_3$ refer to the same person because their ages are the same and Joe is an abbreviation of Joseph. (In comparison, $e_4$ and $e_5$ are not the same person because they have very different ages.) In addition, $e_6$ has an unusually-high age, which can be viewed as an incorrect value. Hence, cleaning this data may involve merging $e_2$ and $e_3$ to a single example $e_{23}$ and fixing or removing $e_6$'s age. However, there are also potential fairness and security issues as we will see in the next sections.

\begin{table}[t]
\centering
\begin{tabular}{| c | c | c | c | c | c |}
\hline
{\bf ID} & {\bf Weight} & {\bf Name} & {\bf Gender} & {\bf Age} & {\bf Label} \\\hline \hline
$e_1$ & 1.0 & John & M & 20 & 1  \\\hline
$e_2$ & 1.0 & Joe & M & 20 & 0 \\\hline
$e_3$ & 1.0 & Joseph & M & 20 & 0 \\\hline
$e_4$ & 1.0 & Sally & F & 30 & 1  \\\hline
$e_5$ & 1.0 & Sally & F & 40 & 0  \\\hline
$e_6$ & 1.0 & Sally & F & 300 & 1 \\\hline
\end{tabular}
\caption{An initial set of training examples with features for predicting whether a person will have high income. The data is not clean ($e_2$ and $e_3$ are duplicates), which may introduce bias that affects model fairness. In addition, $e_6$ has an anomalous age.}
\vspace{-0.3cm}
\label{tbl:examples}
\end{table}


\subsection{Unfairness Mitigation} The machine learning community has been exploring the issue of model fairness where a model may inappropriately discriminate certain demographics. Many definitions of fairness exist (see a recent survey~\cite{DBLP:journals/corr/abs-1810-08810}) and can be largely classified into two categories: local and global. Local measures apply on individuals and make sure the model predictions on them are similar to the predictions on close neighbors. Global measures compare sensitive groups of individuals (e.g., men versus women) and makes sure they have similar statistics. For example, the demographic parity measure dictates that the ratios of positive predictions per group should be the same. Model unfairness is largely due to the inherent bias in the data, which can reflect the opinions of the people who made it and how the data was collected.


More recently, there are mitigation techniques for fixing unfairness, which can be done before (pre-processing), during (in-processing), or after (post-processing) model training~\cite{DBLP:journals/corr/abs-1810-01943}. These techniques typically tradeoff some model accuracy in order to improve model fairness. Among them, we focus on the pre-processing approach where the example weights are adjusted (i.e., reweighed~\cite{DBLP:journals/corr/abs-1810-08810}) to maximize fairness. For example, a simplified reweighing technique for demographic parity is to increase the weights of positively-labeled examples in sensitive groups whose ratio of weighted positive labels is lower than other groups. In Table~\ref{tbl:examples}, the sensitive groups $Gender = M$ and $Gender = F$ have ratios of $\frac{1.0}{1.0+1.0+1.0} = 0.33$ and $\frac{1.0+1.0}{1.0+1.0+1.0} = 0.67$, so we can increase the weight of $e_1$ (the only example that has a positive label in $Gender = M$) from 1.0 to 4.0 so that the ratio is $\frac{4.0}{4.0+1.0+1.0} = 0.67$.

To extend data cleaning to unfairness mitigation, we must understand their possible interactions. In Table~\ref{tbl:examples}, suppose that $e_2$ and $e_3$ are merged into $e_{23}$, which has a total weight of 1.0 and a label of 0. Then the ratio of positive examples in $Gender = M$ increases to $\frac{1.0}{1.0+1.0} = 0.5$, so $e_1$'s weight now needs to be changed from 1.0 to 2.0 (notice this value is smaller than the reweighed result of the previous paragraph) so that $\frac{2.0}{2.0+1.0} = 0.67$.


\subsection{Data Sanitization} The machine learning and security communities are actively studying the problem of robust machine learning against adversarial data in critical applications including spam filtering, autonomous driving, and cybersecurity. A major problem is that the training data is often collected from external data sources, which are vulnerable to attacks by malicious actors~\cite{DBLP:journals/corr/abs-1811-00741}. A popular solution is to make the model training more robust. Another approach that is gaining interest is sanitizing the poisoned data before it is used in training. Data poisoning attacks have recently become more sophisticated~\cite{DBLP:journals/corr/abs-1811-00741}, and there is an arms race on developing better defenses to stop them as well. Data poisoning can also be done on the test data where the same sanitization techniques can apply.

Data sanitization may conflict with data cleaning. For example, consider $e_6$ in Table~\ref{tbl:examples}, which shows an unusually-high age for Sally. This example can be a real attack to confuse the model training, or just an honest typo. Depending on the action taken (e.g., remove $e_6$ or fix its age to a reasonable value), the model training is affected differently as well. In general, using data cleaning and sanitization together can be tricky because incorrect data and poisoned values may be hard to distinguish. While obviously anomalous values may indicate an attack, more recent attacks can generate data closer to the data's distribution and are thus indistinguishable. However, data poisoning is clearly designed to reduce model accuracy while dirty data may have mixed results. Hence, one way to tell the two apart is to see how drastically the model's accuracy changes.

\section{MLClean}
\label{sec:architecture}

Since data cleaning, unfairness mitigation, and data sanitization are ultimately preprocessing the same dataset, it makes sense to unify them. The na\"ive approach of applying each technique independently in any sequence can be problematic for several reasons. Simply ignoring the dependencies between preprocessing techniques may result in incorrect results. For example, if we reweigh examples and then attempt to remove duplicates, then the reweighing may need to be done again to ensure fairness. Moreover, running one operation at a time may have efficiency issues due to redundant operations on the data.

\subsection{Basic Architecture}

\begin{figure}[t]
  \centering
     \includegraphics[width=1.0\columnwidth]{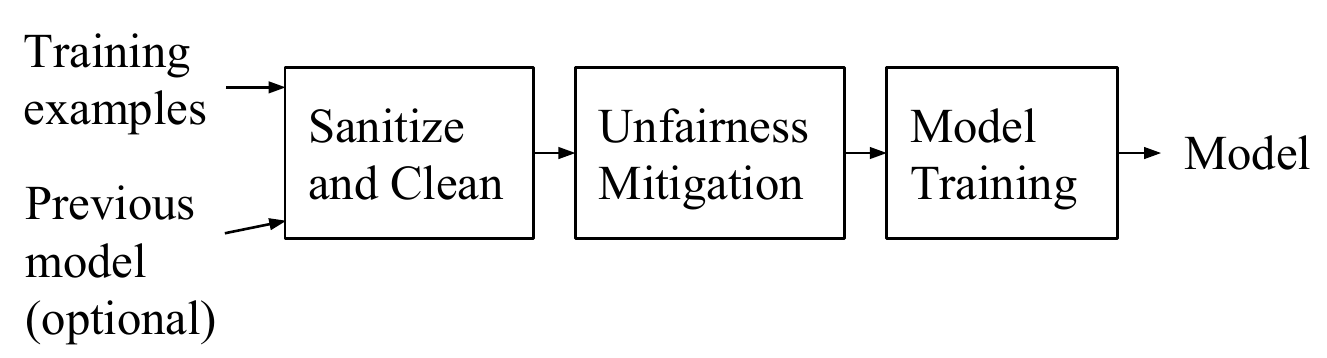}
     \vspace{-0.3cm}
     \caption{The \mc{} architecture. Data sanitization and cleaning are closely related (e.g., anomaly detection followed by entity resolution) and can be done together while unfairness mitigation (e.g., reweighing) is better done afterwards. A previous model is only needed if any of the preprocessing techniques use it.}
 \label{fig:architecture}
 \vspace{-0.4cm}
\end{figure}

We present \mc{}, an extended data cleaning framework that takes into account the dependencies of the three preprocessing techniques and integrates them to produce clean, unbiased, and sanitized data (see architecture in Figure~\ref{fig:architecture}). Data sanitization can be viewed as an extreme version of data cleaning and thus be executed together in one component. The unfairness mitigation component comes afterwards because, while data sanitization and cleaning may affect the bias of data, reweighing examples only changes the example weights and does not effect the correctness of sanitization and cleaning on the other features.

\paragraph*{Example} Suppose the data sanitization is anomaly detection based on clustering and data cleaning is entity resolution~\cite{DBLP:conf/sigmod/ChuIKW16}. We can tightly couple the two techniques by clustering the examples and performing anomaly detection, then running entity resolution on each cluster, assuming they contain the candidate duplicates. This approach is common in entity resolution where candidate duplicates are narrowed down to reduce expensive comparisons. Say we use demographic parity as the fairness measure.

\begin{figure}[t]
  \centering
     \includegraphics[width=0.7\columnwidth]{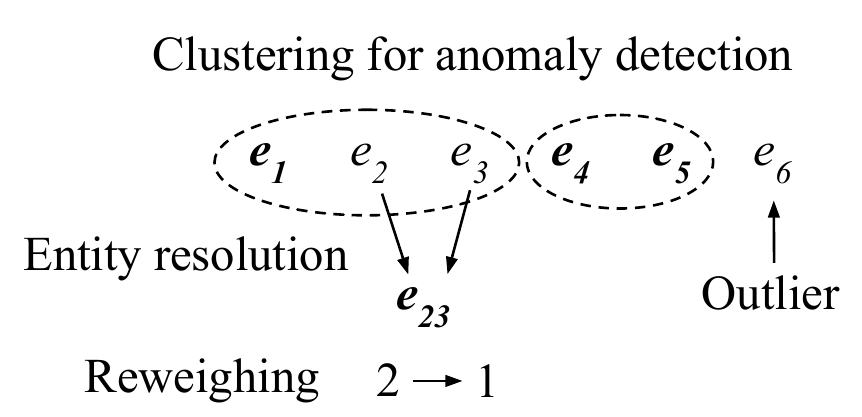}
      \vspace{-0.2cm}
     \caption{\mc{} in action for the examples in Table~\ref{tbl:examples}.}
 \label{fig:examplerun}
 \vspace{-0.3cm}
\end{figure}

Figure~\ref{fig:examplerun} shows how the examples in Table~\ref{tbl:examples} can be preprocessed by \mc{}. First, anomaly detection is used to form the clusters \{$e_1$, $e_2$, $e_3$\} and \{$e_4$, $e_5$\} where $e_6$ is considered an outlier and removed. Suppose an entity resolution process merges $e_2$ and $e_3$ into $e_{23}$ with a summed weight of 2 and a label of 0. Finally, unfairness mitigation adjusts $e_{23}$'s weight from 2 to 1 so that the ratio of positive examples for \{$e_1$, $e_{23}$\} (men) and \{$e_4$, $e_5$\} (women) are the same $\frac{1.0}{1.0+1.0} = 0.5$.

\subsection{Extensions}

The \mc{} framework can be extended to more general combinations of data sanitization, data cleaning, and unfairness mitigation. Beyond removing duplicates, data cleaning can be any general process like HoloClean~\cite{DBLP:journals/pvldb/RekatsinasCIR17}. Data sanitization can also employ more sophisticated defenses~\cite{DBLP:journals/corr/abs-1811-00741}. Of course, one should carefully analyze the possible interactions between each cleaning and sanitization combination.

Some preprocessing techniques may use a previously-trained model as shown in Figure~\ref{fig:architecture}. For example, certain model fairness measures are computed using model predictions. In addition, recall the discussion that data sanitization and cleaning are sometimes hard to distinguish because an example may be adversary or contain an honest typo. In order to distinguish the two cases, we may want to use a previous model and techniques such as influence functions~\cite{DBLP:conf/icml/KohL17} to estimate whether the model's accuracy drastically changes when the example is removed. We believe \mc{} opens up many avenues of research on how to tightly integrate different preprocessing techniques.

\section{Preliminary Results}
\label{sec:experiments}

We evaluate \mc{} and compare it with other baseline approaches. We use the Census Income ($\mathcal{C}$) and German Credit ($\mathcal{G}$) datasets~\cite{DBLP:journals/corr/abs-1810-01943}, which are widely used in the machine learning community. We train a linear model, although \mc{} supports any type of model. To measure model accuracy, we compute the portion of examples that were correctly classified. This metric can be used both for evaluating data cleaning and data sanitization. To measure fairness, we use demographic parity and compute the ratio of the probabilities of positive predictions between two sensitive groups (male/female for $\mathcal{C}$ and old/young for $\mathcal{G}$) as in~\cite{DBLP:journals/corr/abs-1810-01943}. The closer the ratio is to 1, the fairer the model.


For data sanitization, we add adversarial examples to the datasets by using Lasso classifier poisoning~\cite{DBLP:journals/corr/abs-1802-03041} and then defend the attack with clustering-based anomaly detection using the $k$-means algorithm. Here the scenario is that a malicious actor is intentionally poisoning the training data at some external data source. For data cleaning, we introduce duplicates in the datasets by replicating the examples where the degree of replication follows a Zipfian distribution. We then run entity resolution using pairwise comparisons. For unfairness mitigation, we use example reweighing for improving demographic parity as in~\cite{DBLP:journals/corr/abs-1810-01943}.






Table~\ref{tbl:combinations} compares the model accuracy and fairness results of different scenarios. The baseline is when none of the techniques are used (denoted as None). When using the preprocessing techniques individually, $S$ and $C$ improve accuracy while $M$ improves fairness. When running all three techniques in a sequence one by one, the accuracy and fairness results are better overall, although they are not the highest compared to the individual results. In addition, the ordering of the sequence matters. In case of $\mathcal{C}$, running the sequence $\langle M,S,C \rangle$ results in lower fairness than $\langle S,C,M \rangle$ because there is a dependency from $S$ and $C$ to $M$. For $\mathcal{G}$, there is less of a difference because $\mathcal{G}$ was already producing fair models, so there was little unfairness to mitigate in the first place. Finally, we see that \mc{} has similar accuracy and fairness results as $\langle S,C,M \rangle$, but runs significantly faster (by 2.2-2.5x) because it tightly couples $S$ and $C$ where $C$ only runs on clusters identified by $S$. Hence, \mc{} has the best balance of accuracy, fairness, and runtime.




\begin{table}[t]
\centering
\begin{tabularx}{\linewidth}{ |c| *{6}{Y|} }
\hline
   \multicolumn{1}{|c|}{} 
 & \multicolumn{2}{c|}{\bf Accuracy}  
 & \multicolumn{2}{c|}{\bf Fairness}
 & \multicolumn{2}{c|}{\bf Runtime(s)}\\
\cline{2-7}
 {\bf Method} & $\mathcal{C}$ & $\mathcal{G}$ & $\mathcal{C}$ & $\mathcal{G}$ & $\mathcal{C}$ & $\mathcal{G}$\\
\hline \hline
 {\sf None}  & 0.719 & 0.628 & 0.048 & 0.650 & n/a & n/a \\\hline
 $S$  & 0.791 & 0.667 & 0.095 & 0.764 & 0.48 & 0.34 \\\hline
 $C$  & 0.719 & 0.585 & 0.048 & 0.718 & 457.45 & 25.82 \\\hline
 $M$  & 0.731 & 0.614 & 0.400 & 1.005 & 0.02 & 0.02 \\\hline
 $\langle M,S,C \rangle$ & 0.786 & 0.662 & 0.375 & 0.831 & 384.76 & 20.11 \\\hline
 $\langle S,C,M \rangle$ & 0.780 & 0.657 & 0.684 & 0.874 & 391.64 & 19.30 \\\hline
\mc{} & 0.780 & 0.657 & 0.684 & 0.874 & 178.80 & 7.72 \\\hline
\end{tabularx}
\caption{A comparison of running data sanitization ($S$), data cleaning ($C$), and unfairness mitigation ($M$) individually and together on the two datasets.}
\label{tbl:combinations}
\vspace{-0.3cm}
\end{table}


\section{Discussion}
\label{sec:discussion}

In this paper, we contend that data cleaning must be extended to incorporate unfairness mitigation and data sanitization to address the pressing needs of machine learning. We proposed \mc{}, the first framework to tightly integrate data cleaning, unfairness mitigation, and data sanitization. We argue that fixing the source of error (i.e., the data itself) is the best solution for ensuring model accuracy and fairness and demonstrated how the three preprocessing techniques can be used together effectively.

This work is part of a larger trend of Big data - AI integration and opens up many new data management challenges. In particular, we are interested in exploring other combinations of data preprocessing techniques, exploiting white-box models, evaluating \mc{} on general benchmarks, and addressing scalability issues on larger datasets with tighter integrations of preprocessing techniques.

\section*{Acknowledgment}

This research was supported by a Google AI Focused Research Award and the Engineering Research Center Program through the National Research Foundation of Korea funded by the Korean Government MSIT (NRF-2018R1A5A1059921).

\bibliographystyle{ACM-Reference-Format}

\end{document}